\documentstyle[aps,prb,psfig]{revtex}

\begin{document}


\title{Separation quality of a geometric ratchet}

\author{C. Keller, Florian Marquardt, and C. Bruder\\
Departement Physik und Astronomie, Universit\"at Basel,
 Klingelbergstr. 82,\\ CH-4056 Basel, Switzerland}


\maketitle
\begin{abstract}
We consider an experimentally relevant model of a geometric ratchet in
which particles undergo drift and diffusive motion in a
two-dimensional periodic array of obstacles, and which is used for the
continuous separation of particles subject to different forces. The
macroscopic drift velocity and diffusion tensor are calculated by a
Monte-Carlo simulation and by a master-equation approach, using the
correponding microscopic quantities and the shape of the obstacles as
input. We define a measure of separation quality and investigate its
dependence on the applied force and the shape of the obstacles.
\end{abstract}

\section{Introduction}

\begin{figure}
\centerline{\psfig{figure=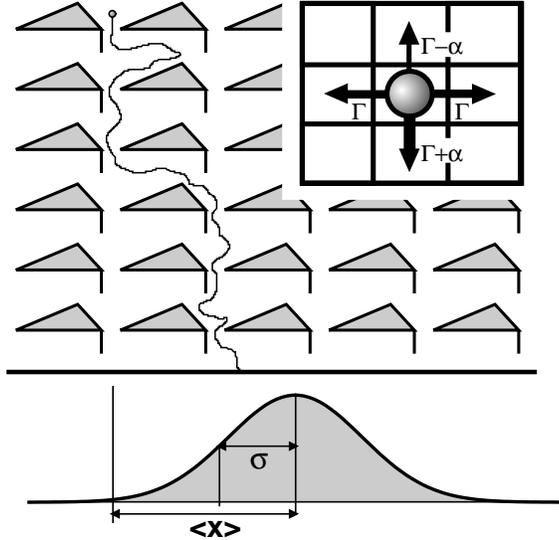,width=8cm}}

\caption{\label{fig1}The model situation: A particle diffusing and
drifting through a periodic array of obstacles. When it has reached
the final row of the array, its horizontal deflection is registered,
which leads to a Gaussian distribution for an ensemble of many
particles (after coarse-graining over several obstacles). The inset
shows the transition rates used in the Monte-Carlo simulations on a
lattice.}
\end{figure}

During the last decade, ratchets have been the subject of intense
research efforts (for a recent comprehensive review see
\cite{reviews}). Ratchets are able to produce a directed current of
particles although no net average force is applied. Besides the
fundamental interest in such a somewhat counterintuitive physical
phenomenon, their analysis is important both for the description of
natural nonequilibrium transport processes (like ``Brownian
motors'' in cells \cite{astumianmolmotor,brownianmotorscience}) and
for concrete technical applications as rectifiers and separation
devices. The various types of ratchets considered so far include
rocking, flashing and correlation ratchets, where a temporally
periodic force, periodic switching of a potential and colored
non-thermal noise, respectively, induce directed transport in an
asymmetric potential \cite{fluxonpump,faucheux,haenggi,reviews}. 
Apart from these, there are
``geometric'' ratchets, which do not necessarily require any
time-dependent forcing but consist, instead, of a two-dimensional
periodic array of asymmetric obstacles 
\cite{duke,ertas,derenyi,oudenaarden,kosterlong,kosturshort}, 
see Fig.~\ref{fig1}. Particles
are driven by a constant external force through the array while they
are undergoing diffusive motion. Because of the asymmetry of the
obstacles, the particles' average drift velocity acquires a component
perpendicular to the direction of the external force, which
constitutes the ratchet effect. Since this is easier to realize
experimentally than the time-dependent ratchets, it has been proposed
and already demonstrated \cite{oudenaarden} as a device for the
separation of charged biomolecules which were subject to an external
electric field and underwent diffusive motion in an array of
micrometer-sized obstacles produced by a lithographic process.

In this article, we want to take up the latter example and analyze
specifically the quality of the separation effect and its dependence
on various parameters. A similar numerical analysis has been carried
out in Ref. \onlinecite{kosterlong}, where the ratchet effect was
investigated for a smooth periodic potential. However, we will
emphasize that optimizing the ratchet effect alone is not equivalent
to optimization of the separation quality. Apart from that, we will
discuss the criterion for assessment of the separation quality, point
out several ``trivial'' possibilities for optimization and analyze
the effective change in the diffusion tensor brought about by the
presence of the obstacles. The discussion will center around a model
of particles undergoing drift-diffusive motion on a discrete
lattice. This model is analyzed numerically using both a
Monte-Carlo scheme as well as the numerical solutions
of a master equation.

The remainder of the article is organized as follows: First, we will
review briefly the geometric ratchet used for the separation of
particles, point out the distinction between microscopic and
macroscopic drift velocities and diffusion tensors and establish a
measure of separation quality. Then, we will set up our lattice
model. Using the results of the numerical simulations, we will present
the dependence of the macroscopic diffusion tensor and drift velocity
on the parameter characterizing the force applied to the particles, as
well as on the shape of the obstacles. Finally, we will optimize the
separation quality for a situation of two particle species that are
subject to different forces, for a restricted set of obstacle shapes.

\section{Separation in a geometric ratchet}

In a ``geometric ratchet'', particles drift and diffuse in a periodic
potential, where the potential inside each elementary cell is
asymmetric, leading to a ratchet effect. In the present work, we will
only consider the special case of a two-dimensional array of
impenetrable obstacles. The drift velocity and diffusion constant of
the particles are assumed to be the same everywhere outside the
obstacles. While the diffusion constant is fixed, the drift velocity
depends not only on the mobility but also on the force applied to the
particles. If the latter derives from an electric field, it will be
proportional to the charge of the particles, which is important for
effecting a separation of different particle species, in the way it
has been demonstrated in Ref. \onlinecite{oudenaarden}. In this work,
differently charged biomolecules have been injected at a corner of a
periodic array of obstacles. An electric field is applied to the setup
so that the particles are subject to a force pointing downward. Due to
the asymmetry of the obstacles, the average drift velocity has a
horizontal component. Therefore, after the particles have traversed
several rows of obstacles, the center of the particle distribution is
deflected by a certain amount. When the magnitude of this deflection
is sufficiently different for two particle species, they may be
separated by collecting the particles arriving at a certain row below
the injection point. This permits a continuous separation of
particles, in contrast to electrophoresis. The quality of separation
does not depend on the strength of the ratchet effect alone but rather
on the difference in deflections for two given species.

\subsection{Macroscopic drift and diffusion}

If the center of the particle distribution moves at a ``macroscopic''
average drift velocity \( \bar{\vec{v}} \), then the slope of the line
it traces (starting from the point of injection) is given by the ratio
\( \bar{v}_{x}/\bar{v}_{y} \). The average deflection in the final
row is obtained by multiplying this ratio with the height of the
array. Obviously, if the average drift velocity vector for one of the
species is just proportional to that of the other one, no separation
can result, regardless of the strength of the ratchet effect
(i.e., the magnitude of the slope) itself.

Apart from the average drift, the particle distribution will also
undergo diffusion, with a ``macroscopic diffusion tensor'' \( \bar{D}
\) that will be different from the microscopic one, due to the
presence of the obstacles (which may hinder the expansion of the
particle cloud in some directions, for example). It is important to
know about the diffusive spreading of the distribution, since it
affects the separation quality. A large difference in deflections for
the two species will be useless if it is bought at the price of a
large width of the respective distributions, which will overlap so
that they cannot be separated unambiguously.

The general functional dependence of \( \bar{\vec{v}} \) and \(
\bar{D} \) can be obtained from dimensional analysis. The microscopic
parameters entering are the microscopic drift velocity \( v \)
(pointing along the \( y \)-direction), the (isotropic) diffusion
constant \( D \), the height of the obstacles \( h \), and a
collection of parameters describing their shape
\( S \) (including the aspect ratio). The only possibility of forming
a dimensionless parameter out of a combination of \( D,\, v \) and \(
h \) is given by

\begin{equation}
\label{xidef}
\xi \equiv \frac{vh}{D}\, .
\end{equation}

It is proportional to the microscopic drift velocity \( v \), and
therefore to the microscopic mobility \( \mu \) multiplied by the
force \( F \). The parameter \( \xi \) will be used to present the
results of our numerical simulations of a lattice model and to compare
them with the real physical parameters. The most general form of
macroscopic drift velocity and diffusion tensor is given by:

\begin{eqnarray}
\bar{\vec{v}} & = & v\, \vec{v}_{0}(\xi ,S)\nonumber \\
\bar{D} & = & D\, D_{0}(\xi ,S)\, .
\label{scaling} 
\end{eqnarray}

Here, \( \vec{v}_{0} \) and \( D_{0} \) are dimensionless vector and
tensor functions.

From these considerations, one can already conclude that two species
differing only in their mobilities (but not in the forces acting on
them) will not become separated, in spite of the ratchet effect. This
is due to the Einstein relation \( D=\mu k_{B}T \), which ensures that
the ratio \( \mu /D \) will be the same for both species. Since the
forces are assumed to be the same, \( v/D \) is also unchanged, such
that the respective values of \( \xi \) are equal. Therefore, the
average drift velocity vector only gets scaled when one passes from
one species to the other, so that the drift slope remains the same, as
has been discussed above.

For small values of the external force (i.e., \( \xi \)), the
macroscopic drift velocity depends linearly on the force and,
therefore, on \( v \) (as long as the linear component is not
suppressed due to symmetry). Therefore, a ``macroscopic mobility
tensor'' \( \bar{\mu } \) relating \( \bar{\vec{v}} \) to the force \(
\vec{F} \) may be defined, such that \( \bar{\vec{v}}=\left( \bar{\mu
}/\mu \right) v\vec{n} \), with \( \vec{n} \) the direction of the
microscopic drift. The numerical calculations (see remarks in section
\ref{latticemodel}) will confirm that it fulfills the Onsager symmetry
relations: \( \bar{\mu } \) is found to be symmetric. Likewise the
macroscopic diffusion and mobility tensors are connected by the
Einstein relation for small values of the external force: \(
\bar{D}=\bar{\mu }k_{B}T \), where the temperature \( T \) is obtained
from the ratio \( D/\mu \) of the corresponding microscopic
quantities. Physically, this derives from the fact that the
equilibrium distribution in a setup with a wall at the bottom is given
by the Boltzmann distribution, which carries over from the microscopic
density to the coarse-grained density, whose evolution is governed by
\( \bar{D} \) and \( \bar{\mu } \). Apart from the fact that one has
to be still in the regime of linear variation of \( \bar{\vec{v}} \)
for the Einstein relation to make sense, it can only hold as long as
the force is not so strong as to make the density fall off rapidly
over the scale of a single obstacle, because then the coarse-graining
procedure is no longer justified. This happens approximately at \( \xi
\sim 1 \), which is also the condition that has to be reached to see
an appreciable separation effect.

\subsection{Analytical estimate}

\begin{figure}
\centerline{\psfig{figure=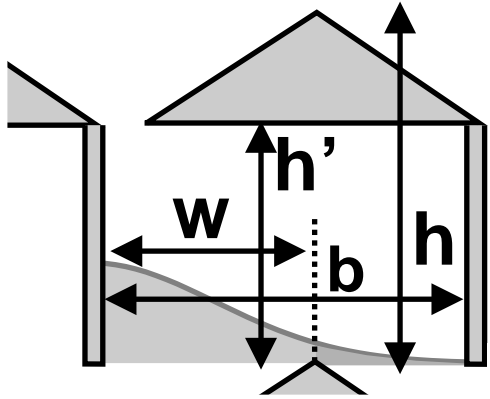,width=8cm}}
\caption{Concerning the analytical estimate of the
macroscopic drift velocity: The spreading Gaussian distribution is cut
in two halves (see text). }
\label{cutit}
\end{figure}

The magnitude of the ratchet effect can be estimated analytically
\cite{faucheux,duke,derenyi} for high enough external
force (large \( \xi \)). Then, one can treat the motion in the
direction of the external force as deterministic (neglecting
diffusion), so that diffusive spreading takes place only in the
perpendicular direction. In this simplified picture, the geometric
ratchet becomes analogous to a time-dependent one-dimensional
``flashing'' ratchet, with the \( y \)-coordinate playing the role
of time. For a typical obstacle (as it has been used, e.g., in the
experiment of van Oudenaarden and Boxer \cite{oudenaarden}), the
diffusing particle distribution (in the shape of half a Gaussian
curve) will be split in two parts by the ``top part'' of the obstacle,
see Fig.~\ref{cutit}. Here and in the following, we assume that there
is one connected obstacle. The left part proceeds downward further on,
while the right part moves one cell to the right. Therefore, the slope
\( \bar{v}_{x}/\bar{v}_{y} \) is given essentially by the percentage
of particles that have moved to the right in such an ``elementary
step'', i.e., by an integral over the respective part of the
Gaussian distribution. For large \( \xi \) it can be approximated by
an exponential (which becomes a good approximation if the magnitude of
the following exponent exceeds \( 2 \)):

\begin{equation}
\label{anest}
P(\xi )\equiv \frac{\bar{v}_{x}}{\bar{v}_{y}}\approx 
\frac{2b}{\sqrt{\pi }w}\sqrt{\frac{h'}{h}}\frac{1}{\sqrt{\xi }}
\exp \left( -\frac{w^{2}}{4hh'}\xi \right) \, .
\end{equation}

Here, the dependence on \( \xi \) has been
made explicit, while all the other dimensionless parameters are ratios
of lengths determining the shape \( S \) of the obstacle (see
Fig.~\ref{cutit}). Obviously, if the force becomes very large, the
particles will only move down inside ``channels'', since they do not
have time to spread to the left or right. Then the slope \( \propto
\bar{v}_{x}/\bar{v}_{y} \) becomes very small, as is expressed by this
formula. On the other hand, for very small forces neglecting the
possibility of diffusing backwards in \( y \)-direction (or more than
one cell in \( x \)-direction) renders this estimate invalid.
Qualitatively, however, it is correct that the slope tends to a
constant in the limit of vanishing force \( \xi \rightarrow 0
\). Therefore, separation is ineffective both at very small forces
(since the slopes of two species are the same) and at very large
forces (since the slopes differ but are both very small). Since the
prefactor multiplying \( \xi \) in the exponent is of the order of
one, the optimum separation quality for the obstacle shape discussed
here will also be reached when \( \xi _{1} \) and \( \xi _{2} \) 
are of the order of one. This is confirmed by the numerical
analysis below.

\subsection{Quality of separation}

In the long-time limit, the ensemble of particles of a given species
(having started at the injection point) assumes the form of a
two-dimensional Gaussian distribution, which drifts at a velocity \(
\bar{\vec{v}} \). Therefore, the distribution of particles along the
\( x \)-coordinate of the final row is also a Gaussian, which is
centered around some value \( \left\langle x\right\rangle \) and has a
width \( \sigma \). We will assume that separation is performed by
collecting all the particles up to some point \( x_{0} \) in one bin
and the rest in another bin. Ideally, the two bins would only receive
particles of a single species (\( 1 \) or \( 2 \)). Due to the overlap
of the two distributions, this is not possible and there is a certain
percentage of particles that are attributed to the wrong
bin. Qualitatively, the optimal choice of \( x_{0} \) is one where
this ``error'' is minimized. However, since there are two different
types of errors (percentage of particles \( 1 \) attributed to bin \(
2 \) and vice versa), no unambiguous definition of the optimal choice
of \( x_{0} \) and the corresponding optimal separation quality
exists. We suggest to take the separation of \( \left\langle
x_{1}\right\rangle \) and \( \left\langle x_{2}\right\rangle \) and
divide it by the maximum of the widths \( \sigma _{1} \) and \(
\sigma _{2} \) to arrive at a measure of separation quality which is
easily evaluated:

\begin{equation}
\label{sepqual2}
Q\equiv \left| \left\langle x_{1}\right\rangle -
\left\langle x_{2}\right\rangle \right| /\max(\sigma _{1},\sigma _{2})
\, .
\end{equation}

 For the parameters
considered here it seems to be appropriate.

Another possible definition consists in replacing the maximum by the 
geometric mean of the two widths:

\begin{equation}
\label{sepquality}
Q'\equiv \left| \left\langle x_{1}\right\rangle -
\left\langle x_{2}\right\rangle \right| /\sqrt{\sigma _{1}\sigma _{2}}
\, .
\end{equation}

Note, however, that there are situations when this may be a misleading
measure, particularly when one of the widths \( \sigma _{1} \) or \(
\sigma _{2} \) is much larger than the other. In these cases, $Q$ will probably
be better suited.

Other measures have been used in the literature. For example, in
Refs. \onlinecite{derenyi,duke}, the authors essentially asked
how large the relative difference in diffusion constants of two
species should be in order to have a separation that exceeds the
spread of one of the two distributions.

In any case, any reasonably defined optimal separation quality can
only be a function of two dimensionless parameters: the measure \( Q
\) (or $Q'$) used here and the ratio \( \sigma _{1}/\sigma _{2} \).

The goal is to optimize the separation quality for two given species
of particles by varying the applied force and the shape of the
obstacles (including their size and aspect ratio). The following
parameters are naturally assumed to be fixed: the microscopic
mobilities and diffusion constants of the species (connected by the
Einstein relation for a given temperature), the ratio \( \lambda \) of
the forces (equal to the ratio of charges in an electric field) and
the total height \( H \) of the periodic array in \( y
\)-direction. The latter will be dictated by practical considerations
(if it could be made arbitrarily large, one would do this to get an
ideal separation effect).

We want to argue when and why a nontrivial optimum is to be
expected. It has been pointed out above that choosing the force too
large or too small results in a decrease of separation
quality. Likewise, one could make the ratchet effect itself
arbitrarily strong, by choosing a slanted line of large slope \( dx/dy
\) (as one big obstacle). However, both species would drift along that
line and could not be separated either.

In order to obtain a more quantitative understanding, we use the
general functional forms given for \( \bar{\vec{v}} \) and \( \bar{D}
\), and insert them into the measure \( Q \) introduced above. More
precisely, we use the scaling expressions \( \left\langle
x\right\rangle =H\bar{v}_{x}/\bar{v}_{y}=Hx_{0}(\xi ,S) \) and \(
\sigma =\sqrt{DH/v}\sigma _{0}(\xi ,S) \) for both species, which are
assumed to have the same microscopic values of \( \mu \) and \( D \)
but different forces acting on them, such that \( v_{2}=\lambda
v_{1}\equiv \lambda v \) and \( \xi _{2}=\lambda \xi _{1}\equiv
\lambda \xi \). Thus, we obtain:

\begin{eqnarray}
Q & = & \frac{\left| \left\langle x_{1}\right\rangle -
\left\langle x_{2}\right\rangle \right| }{\max(\sigma _{1},\sigma
 _{2})}
\nonumber \\
 & = & \sqrt{\frac{H}{h}}\frac{\sqrt{\xi }\left|
x_{0}(\xi )-x_{0}(\lambda \xi )\right| }{\max(\sigma _{0}
(\lambda \xi )/\sqrt{\lambda}, \sigma _{0}(\xi ))}\, .
\label{Qdimensionless} 
\end{eqnarray}

After the fraction in the last line has been optimized by varying both
\( \xi \) and the shape \( S \) of the obstacles, \( Q \) could be
made arbitrarily large by having the height \( h \) of an individual
obstacle go to zero (at fixed aspect ratio), such that the array (of
fixed height \( H \)) contains more rows. Since \( \xi =vh/D \) must
remain constant, this means that the microscopic drift velocity \( v
\) has to go to infinity. Physically, the enhancement of separation
quality can be understood in the following way: Although the slopes
will remain the same, the relative size of the diffusive spread
decreases like \( \sqrt{h/H} \). Note as well that, due to the same
reason, the maximum of the fraction does not exist, strictly
speaking. It is possible to keep \( h \) fixed but still have an
effective increase in the number of rows by placing an array of
miniaturized obstacles inside an ``elementary'' cell. However, in
practice there are obvious restrictions on the force that can be
applied to the particles, as well as on the minimum size of the
obstacles. Therefore, it is only possible to choose the force as large
as possible and the corresponding value of \( h \) in such a way that
\( \xi \) takes on the optimal value under these restrictions.

It is also helpful to consider the behavior of \( Q \) for the
analytical estimate given in Eq.~(\ref{anest}), which is valid for
large values of \( \xi \). After \( H/h \) rows (and a time \(
t=H/\bar{v}_{y} \)), the average deflection is \(
\bar{v}_{x}t=\bar{v}_{y}Pt \) and the variance of the resulting Poisson
distribution is \( 2\bar{D}_{xx}t=bHP \). This gives the relation
\( \bar{D}_{xx}/\bar{v}_{x}=b/2 \), which is confirmed by the
numerical results below (for obstacles of type ``A'' and ``B''
in Fig. \ref{deflection}). We obtain:

\begin{equation}
Q\sim \sqrt{\frac{H}{b}}\frac{|P(\xi )-
P(\lambda \xi )|}{\max( \sqrt{P(\xi )}, \sqrt{P(\lambda \xi )} ) }\, .
\label{Qestimate}
\end{equation}

By setting \( P(\xi )\sim \exp (-\gamma \xi ) \), with some
exponent \( \gamma \), we can find the approximate behavior of $Q$: 
It drops like \( \exp
(-\xi \gamma /2) \) at large \( \xi \), regardless of \( \lambda (>1)
\). On the other hand, the alternative definition $Q'$ given above would
only decrease like \( \exp (-\xi \gamma
(3-\lambda )/4) \). For \( \lambda >3 \), 
this may even rise at large \( \xi \), which is probably an indication
that, for this regime,
the expression for \( Q' \) is not any longer a good
measure of separation quality: The difference in spread of the two
species grows too fast.

\section{The lattice model}

\label{latticemodel}

We have set up a model to study numerically the diffusion of particles
under the influence of an external force in a periodic array of
obstacles: A single particle is positioned on a point of a
two-dimensional lattice which consists of square fields. During each
time-step, the particle either changes its position to one of the
neighboring squares (with a certain probability) or it remains on its
original square.

The probabilities to move to the right or to the left are both equal
to \( \Gamma \). In the absence of obstacles, there is consequently
no net flow of particles in the horizontal direction. The probability
to move downward (i.e. in positive \( y \)-direction) is \( \Gamma
+\alpha \), the probability to move upward is \( \Gamma -\alpha \)
(see inset of Fig. \ref{fig1}). The probability to remain on the
original square is therefore equal to \( 1-4\Gamma \). Because of the
different probabilities to move up- and downward, a net flow results
in the vertical direction.

The following relations hold for the microscopic drift velocity 
\( v\) in y-direction and the microscopic diffusion constant \( D \):

\begin{eqnarray}
v & = & 2\alpha \nonumber \\
D & = & \Gamma \, .
\label{alphagamma} 
\end{eqnarray}

We assume lengths to be measured in lattice constants and time in
units of the elementary time-step. For the interpretation of the
results, only the value of the dimensionless parameter \( \xi \) is
needed:

\begin{equation}
\label{xiagain}
\xi =vh/D=2\alpha h/\Gamma \, ,
\end{equation}

where \( h \) is the height of an obstacle (elementary cell of the
array) measured in lattice constants. From this expression, it is
clear that increasing the spatial resolution (\( h \)) means
decreasing \( \alpha /\Gamma \), such that this ratio vanishes in the
physically relevant continuum limit. Actually, we have already assumed
\( \alpha \ll \Gamma \) in writing down Eq.~(\ref{alphagamma}).
Otherwise, one would have to take into account that the diffusion
resulting from the model stated above is anisotropic, with \(
D_{xx}=\Gamma \) and \( D_{yy}=\Gamma -2\alpha ^{2} \). In our
Monte-Carlo simulation, we have
usually chosen \( \Gamma \) to be of the order of \( 0.1 \) and \(
\alpha /\Gamma \) to be less than about \( 0.1 \).

Obstacles are represented by ``forbidden'' squares. If the particle
would have to move onto such a square, it does not move and remains on
its original square. All obstacles are arranged periodically.

Particles start in the top left square of the array. They continue
moving until they have reached the final row, where their \( x
\)-coordinate is saved. Using the results of many runs, the average
deflection \( \left\langle x\right\rangle \) and the standard
deviation \( \sigma \) of the resulting distribution can be
calculated, see Fig.~\ref{fig1}. Alternatively, a particle runs for a
given number of time steps and its final coordinates are
registered. In this way, the macroscopic drift velocity \(
\bar{\vec{v}} \) and diffusion tensor \( \bar{D} \) can be obtained.

In order to test the Monte-Carlo simulation (which has been
implemented in C++), we have verified that the distribution of
deflections in the final row is Gaussian (when coarse-grained over
several obstacles) and becomes independent of the precise starting
position after a sufficiently large time (number of rows of the
array), and all quantities show the correct scaling behavior described
by Eq.~(\ref{scaling}), provided \( \alpha /\Gamma \) is chosen small
enough. The lattice resolution (connected with \( \alpha /\Gamma \))
has been chosen such that the results do not depend on it appreciably
any more.

Furthermore, at small values of the external force (i.e., of \( v \)
and \( \xi \)), the macroscopic drift velocity depends linearly on
this force and fulfills the Onsager symmetry relations. Likewise the
Einstein relation holds: \( \bar{D}=\bar{\mu }k_{B}T \).

\section{Master equation}

While every quantity of interest (macroscopic drift velocity and
diffusion tensor, average deflection and spread of distribution) can
be calculated using the Monte-Carlo simulation, it is nevertheless
useful to consider a master equation solution as well. This is both
because reaching a high statistical accuracy requires a large number
of Monte-Carlo runs and because a discussion of the master equation
yields additional physical insights.

The particle distribution \( p(x,y) \) is defined on the lattice, with
integer coordinates \( x \) and \( y \). In a single time-step, the
distribution changes by

\begin{eqnarray}
\delta p(x,y) & = & \Gamma (p(x+1,y)+p(x-1,y))\nonumber \\
 & & +(\Gamma -\alpha )p(x,y+1)+(\Gamma +\alpha )p(x,y-1)\nonumber \\
 & & -4\Gamma p(x,y)\, ,\label{Ldef} 
\end{eqnarray}

where the quantities on the right-hand side are to be evaluated at
time \( t \) and \( \delta p\equiv p(t+1)-p(t) \). The equation shown
here holds for every site \( (x,y) \) which has no neighboring
obstacle sites. For each ``forbidden'' site, the corresponding
incoming and outgoing rates have to be left out. The temporal
continuum limit is obtained by letting \( \Gamma \) and \( \alpha \)
tend to zero, with their ratio kept fixed.

At large times, an ensemble of particles that has started at the
injection point will be spread over many obstacle cells. Viewed on the
scale of only a few cells (much less than the total spread), the
distribution \( p \) is periodic. Therefore, we can calculate the
average drift velocity by solving for the stationary distribution \( p
\) defined inside the cell of a single obstacle, imposing periodic
boundary conditions. \( \bar{\vec{v}} \) is then given by the average
velocity inside the cell, i.e., by

\begin{equation}
\label{driftcurrent}
\bar{\vec{v}}=\sum_{\text{links}}\vec{j}\, .
\end{equation}

Here, the sum runs over all links connecting adjacent sites and the
``current density'' \( \vec{j} \) along each link is obtained by
multiplying the values of \( p(x,y) \) on the two connected sites by
the transition probabilities of a jump along the link, taking into
account the direction of the link and leaving out blocked sites. The
distribution \( p \) itself is assumed to be normalized: \( \sum
p(x,y)=1 \).

Actually, every site with no neighboring obstacle sites contributes
just \( p(x,y)\, \vec{v} \) to this sum, where \( \vec{v} \) is the
microscopic drift velocity. In this sense, the deviation of the
macroscopic drift \( \bar{\vec{v}} \) from \( \vec{v} \) is seen to
arise only from the boundaries of the obstacle. This can be understood
most easily in the continuum limit, where \( \vec{j}=-D\vec{\nabla
}\rho +\vec{v}\rho \) (\( \rho \) being the continuous
distribution). An integration by parts yields:

\begin{equation}
\label{driftcurrent2}
\bar{\vec{v}}=\int _{cell}\vec{j}\, d^{2}\vec{r}=\vec{v}+
D\oint _{obst.}\rho \vec{n}\, ds\, .
\end{equation}

Here, the second integral extends over the boundary of the obstacle and
\( \hat{n} \) is the normal vector of that boundary (pointing away
from the obstacle). Therefore, those boundaries where the density \(
\rho \) ``piles up'' (see Fig.~\ref{densitycurrent} below) contribute
most to the change in drift velocity. The second integral can
therefore be interpreted as being proportional to the ``force'' due to
the obstacle, impeding and deflecting the free flow of particles.

For the numerical solution, we first set up the (sparse) matrix
corresponding to the transition rates on the lattice with periodic
boundary conditions and then solve for the stationary solution \( p
\). This is done by setting \( p(x,y)=1 \) on an arbitrary non-blocked
site and striking out the respective column and row in the homogeneous
linear system of equations, such that it becomes inhomogeneous and
nonsingular. We have used a sparse matrix bi-conjugate-gradient solver
from the LAPACK package of linear algebra routines \cite{lapack}. For
a typical cell size of \( 128\times 128 \), the number of nonvanishing
matrix entries is about \( 10^{5} \). In the end, \( p \) is
normalized and the current density and macroscopic drift velocity are
calculated. This type of calculation has already been performed by
the authors of Ref. \onlinecite{kosterlong}, for a model with a smoothly
varying periodic potential.

It is also possible to calculate the macroscopic diffusion tensor \(
\bar{D} \) using the master equation. This involves the solution of an
inhomogeneous linear equation, with the linear operator defined by
Eq.~(\ref{Ldef}) and the inhomogeneity derived from the solution \( p
\) of the homogeneous equation. The derivation of the equation and the
formula for \( \bar{D} \) can be found in the Appendix. To the best
of our knowledge, such an analysis has not been carried out before.

\section{Numerical results}

In this section, we present the results of both the Monte-Carlo simulations
and solutions of the master equation.

In the Monte-Carlo simulation, the relative statistical accuracy of
the macroscopic drift velocity (\( \delta \bar{v}_{x}/\bar{v}_{x} \)
etc.) is given approximately by \( \sqrt{D/\left( v^{2}Nt\right) } \),
while that of \( \bar{D} \) is estimated to be \( 1/\sqrt{N} \), where
\( N \) is the number of samples and \( t \) the number of
time-steps. Independence of the detailed initial conditions is reached
when the diffusive spread \( \sqrt{Dt} \) becomes much larger than the
width of an obstacle cell (typically \( 160 \)). We have chosen values
of \( t\geq 10^{7} \) and \( N\geq 10^{3} \) in order to fulfill these
criteria.

\subsection{Density distribution and flow field}

\begin{figure}[t]
\centerline{\hbox{\psfig{figure=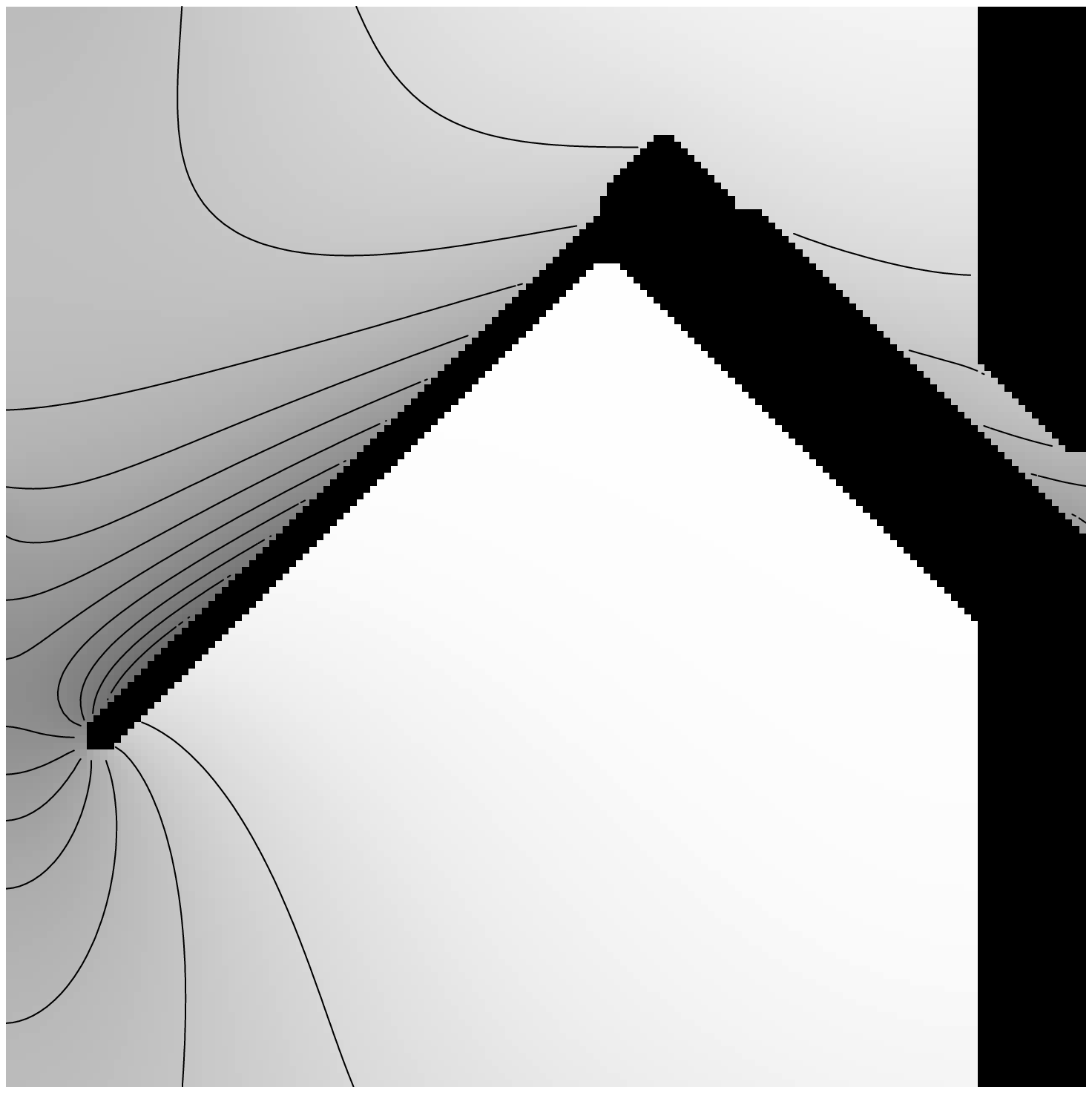,width=6.75cm}\quad
\psfig{figure=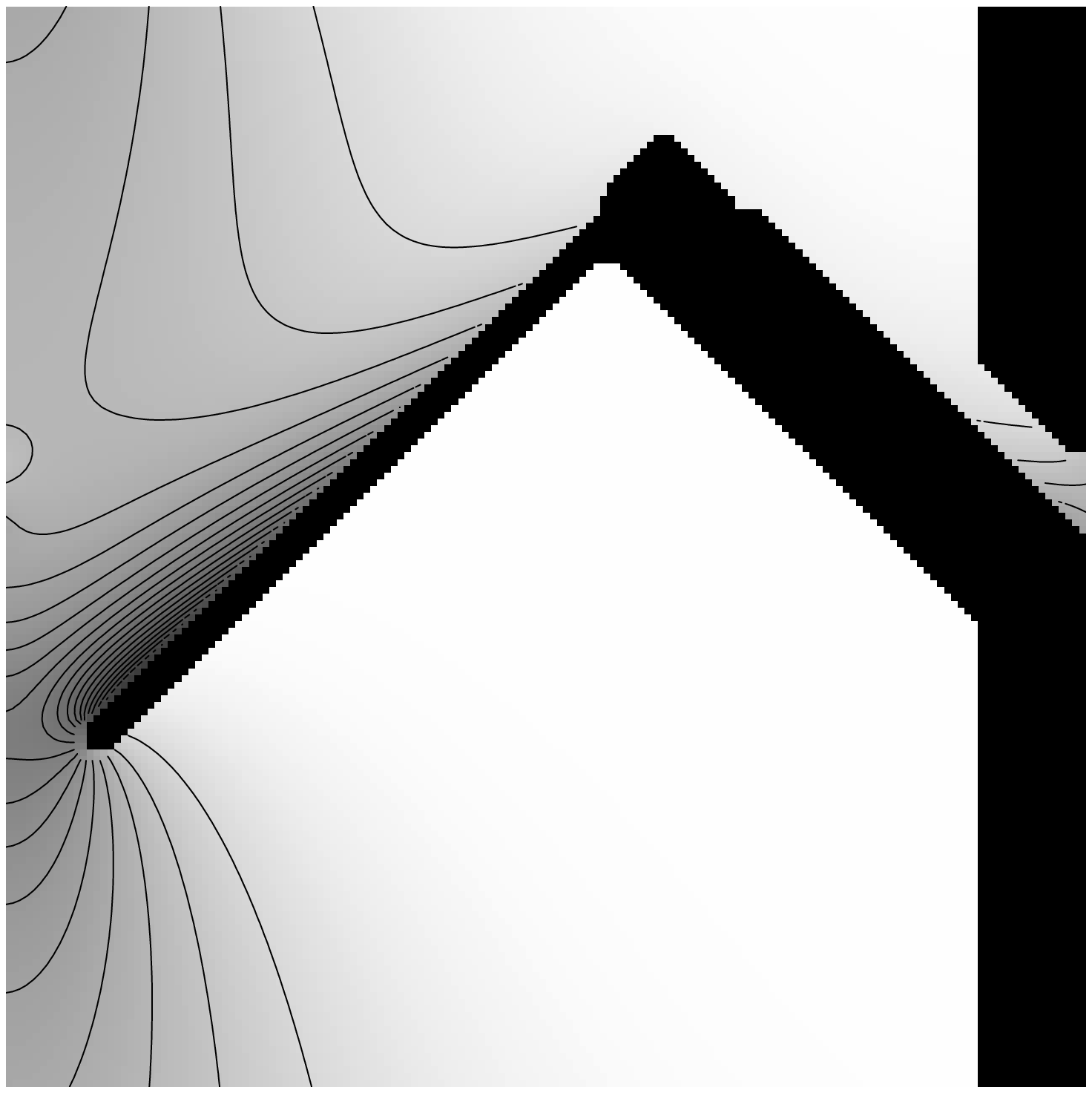,width=6.75cm}}}
\centerline{\hbox{\psfig{figure=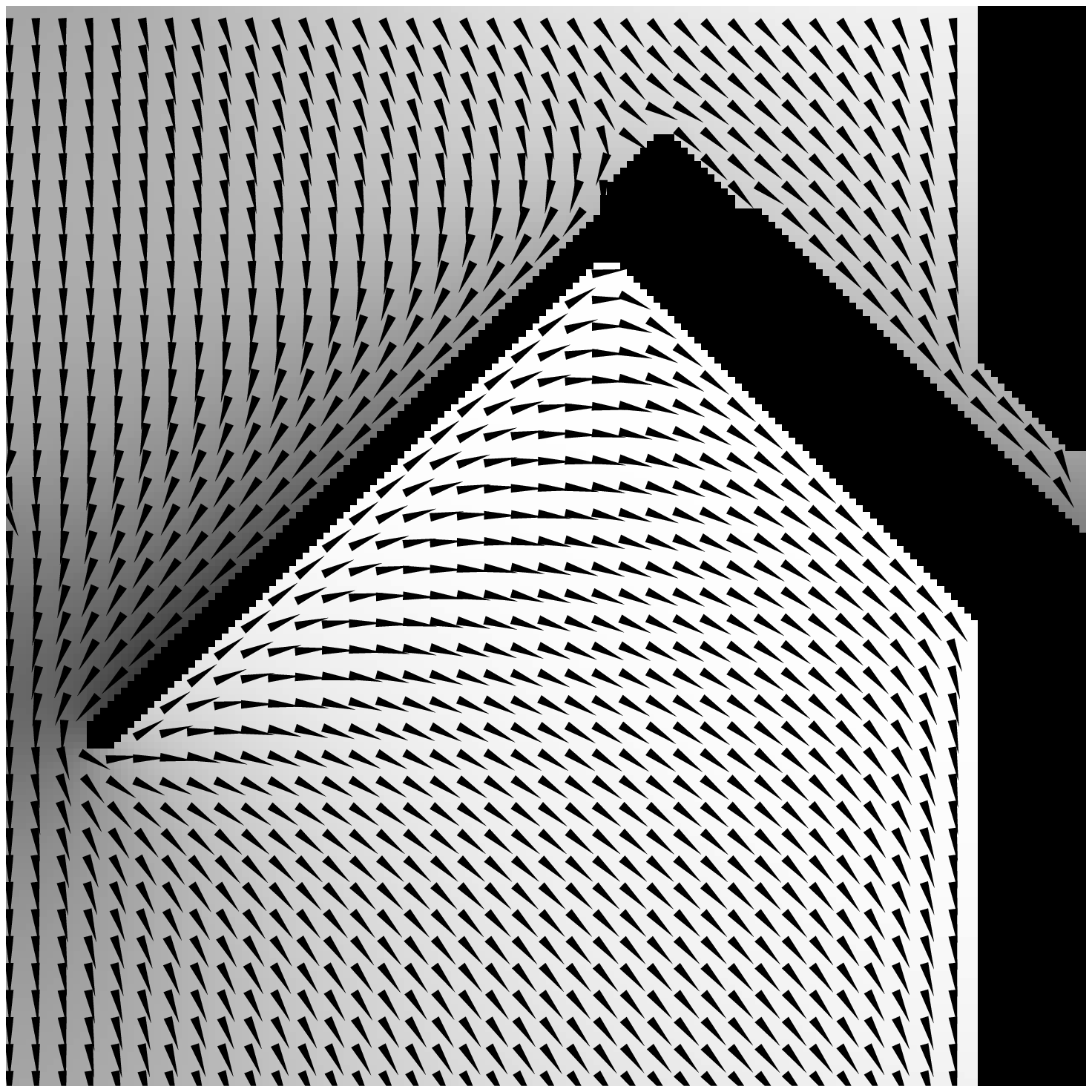,width=6.75cm}\quad
\psfig{figure=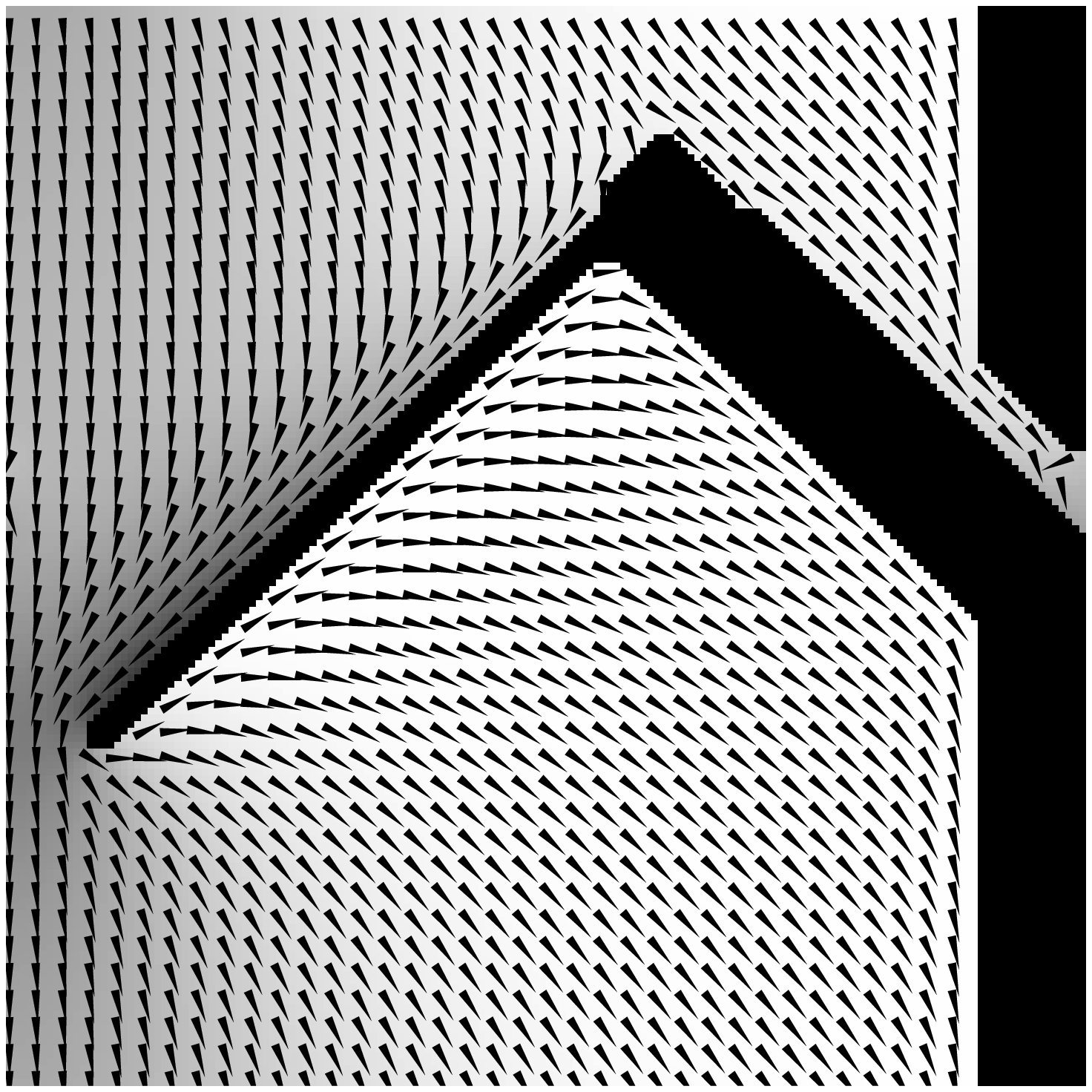,width=6.75cm}}}
\centerline{\hbox{\psfig{figure=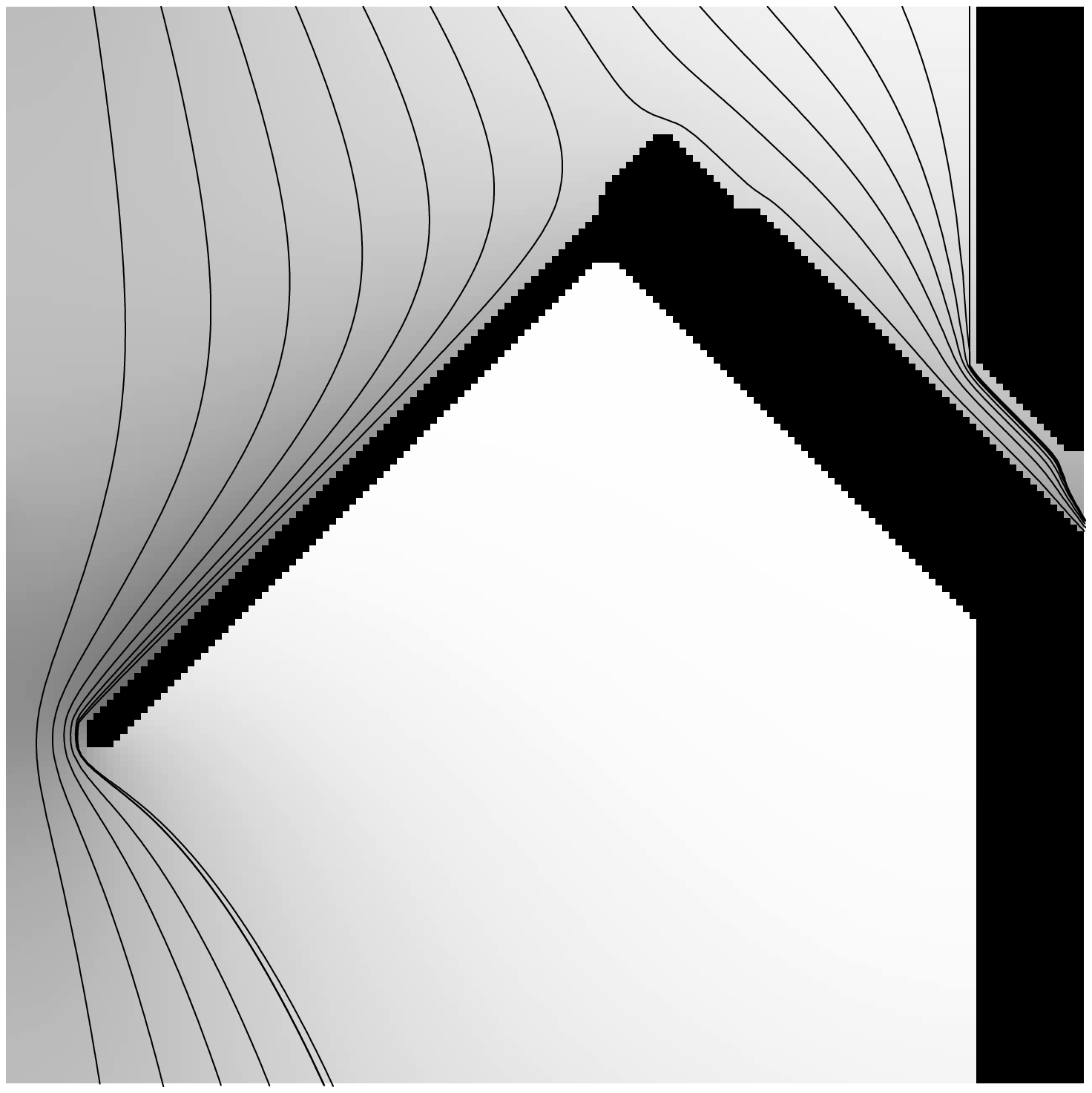,width=6.75cm}\quad
\psfig{figure=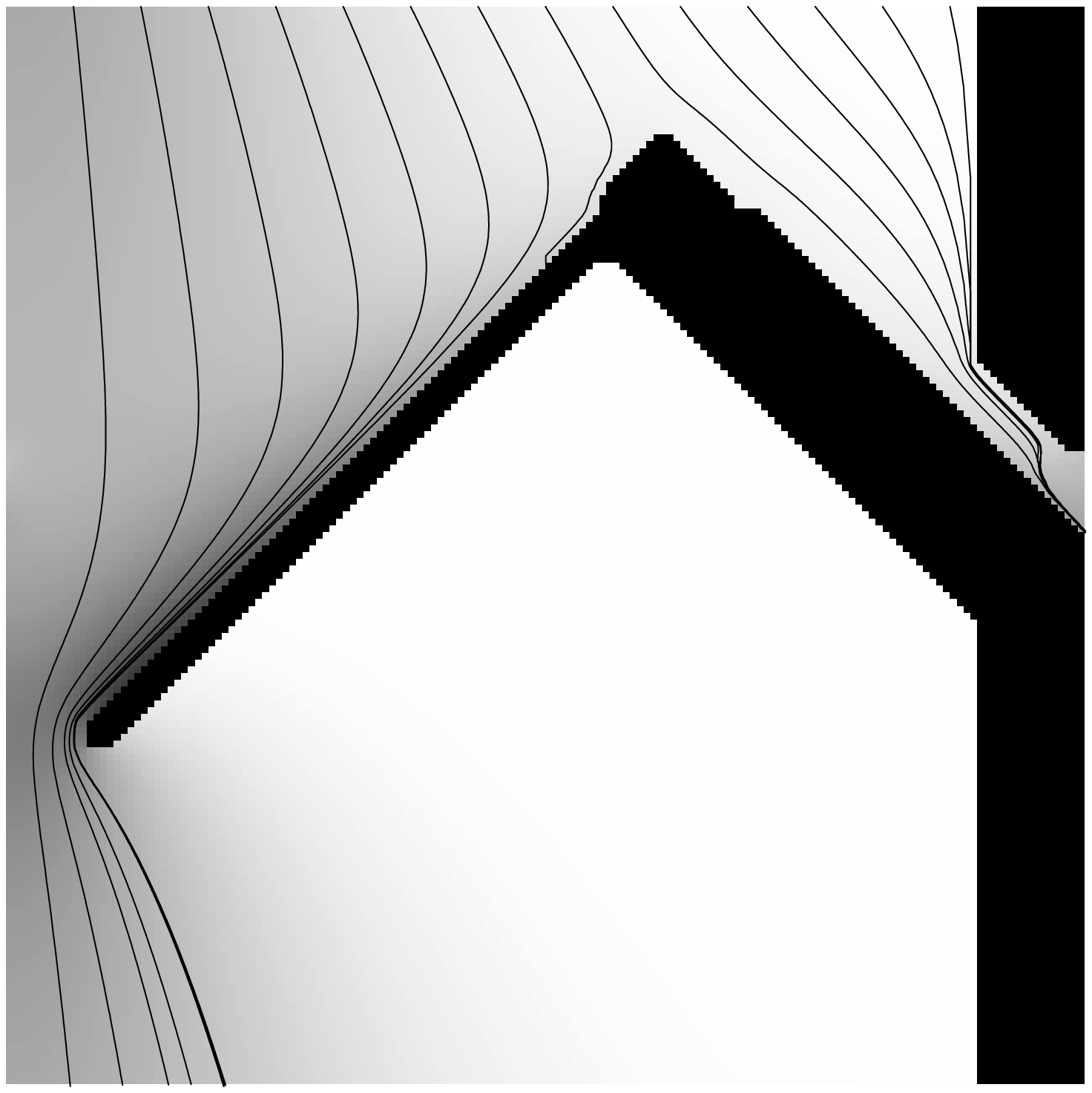,width=6.75cm}}}
\caption{
\label{densitycurrent}
Distribution \( p \) (top), 
velocity direction field \( \hat{v} \) (middle), 
and ``streamlines'' (bottom) for low and high force (left and
right columns). The strength of the force \( \xi =12 \) for the left column
and \( \xi =24 \) for the right column, with the
microscopic drift pointing downward. Darker shading signifies
increased \( p \). Contour lines for \(
p \) have been chosen at the same (equidistant) values for the
two pictures at the top. Note: The streamlines serve
to visualize the direction of the current density, while the real
motion of a single particle is governed by drift and diffusion. The
``velocity direction'' is the normalized current density vector field.
Note the stronger pile-up in density at the left ``roof'' of the obstacle
for the stronger force, as well as the distortion of streamlines there and
in the narrow channel at the right of the pictures. 
The lattice shown here consists of \( 160\times
160 \) sites.}
\end{figure}

The particle distribution \( p \) and corresponding current density \(
\vec{j} \) resulting from the solution of the master equation for a
typical obstacle shape are depicted in Fig.~\ref{densitycurrent}, both
for a low and a high value of \( \xi \). This corresponds to the
optimal situation for the separation of two species, where one of them
is almost not deflected at all (high force, i.e., high \( \xi \)),
while the other one has an appreciable probability to go to a
neighboring cell to the right, due to diffusion around the top part of
the obstacle.

\subsection{Macroscopic drift velocity and diffusion tensor}

\begin{figure}
\centerline{\psfig{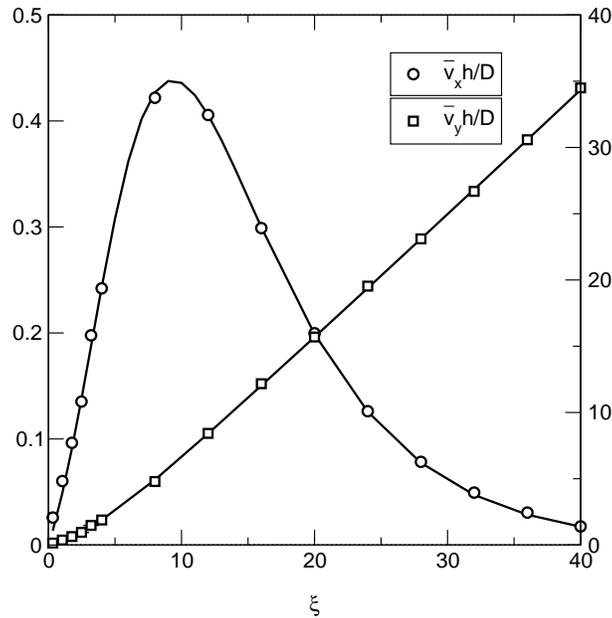}}

\caption{\label{macdrift}Macroscopic drift velocity components
\( \bar{v}_{x} \) (circles, left scale) and \(
\bar{v}_{y} \) (squares, right scale) in units of \(
D/h \), plotted vs. \( \xi =vh/D \), for the
obstacle shape shown in Fig.~\ref{densitycurrent}, with a lattice
resolution of \( 160\times 160 \). The lines indicate
the results of the master equation, the symbols (which are larger than
the error bars) are from Monte-Carlo simulations (\(
t=2\times 10^{7},\, N=10^{3} \)).}
\end{figure}

The results for the components of the macroscopic drift velocity are
displayed in Fig.~\ref{macdrift}. These data have been obtained for
the obstacle of Fig.~\ref{densitycurrent}, using both the master
equation with periodic boundary conditions and the Monte-Carlo
simulation. As has been explained in a previous section, the onset of
a nonlinear dependence of \( \bar{v}_{y} \) and \( \bar{v}_{x} \) on
\( \xi \) at rather low \( \xi \) implies that the obstacle is
well-suited for obtaining a separation effect (rather than merely a
ratchet effect).

\begin{figure}
\centerline{\psfig{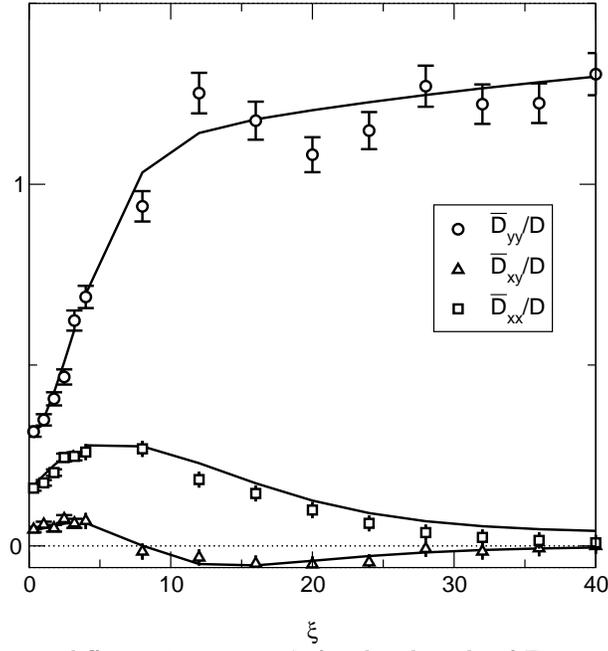}}

\caption{\label{macdiff}Components of macroscopic diffusion tensor
vs. \( \xi \), for the obstacle of
Fig.~\ref{densitycurrent}. Circles: \( \bar{D}_{yy}/D
\), squares: \( \bar{D}_{xx}/D \), triangles:
\( \bar{D}_{xy}/D \). Symbols show results of the
Monte-Carlo simulation, lines those of the master equation. We
attribute the deviation at higher values of \( \xi \)
(and, therefore, \( \alpha /\Gamma \)) to effects of
the finite grid resolution.}
\end{figure}

In Fig.~\ref{macdiff}, the components of the (symmetrized) macroscopic
diffusion tensor \( \bar{D} \) have been plotted vs. \( \xi \), for
the same obstacle. At low values of the driving force, all components
of \( \bar{D} \) are generally reduced compared with the microscopic
diffusion constant \( D \), since the obstacles hinder the free
diffusion of particles. At higher forces, the \( y \)-component is
increased, while the \( x \)-component \( \bar{D}_{xx} \) is further
reduced. This can be understood in the following way: The motion
proceeds inside vertical ``channels'', such that the particle cannot
move easily in horizontal (\( x \)) direction, while the diffusion in
\( y \)-direction is more or less free. Comparing the data for \(
\bar{D}_{xx} \) and \( \bar{v}_{x} \) shows that \(
\bar{D}_{xx}/\bar{v}_{x}\approx b/2 \) at larger values of \( \xi \)
(to the right of the maximum of \( \bar{v}_{x}(\xi ) \)), as expected:
both quantities decrease exponentially. Note that the off-diagonal
component \( \bar{D}_{xy} \) of the macroscopic diffusion tensor
changes sign at about \(\xi\approx 8 \), which seems to approximately 
coincide with the sign-change in \(\nabla\bar{v}_{x}/\nabla\bar{v}_{y}
\) (see Fig.~\ref{macdrift}). We have not come up, however, with an
explanation for the approximate correlation between \( \bar{D}_{xy}/D
\) and \(\nabla\bar{v}_{x}/\nabla\bar{v}_{y} \) yet.

The statistical accuracy of the Monte-Carlo results for \( \bar{D} \)
is worse than that of the \( \bar{v} \) results, as expected. The
deviation between the results of the master equation and Monte-Carlo
simulation at larger values of \( \xi \) is reduced when the grid
resolution is enhanced (i.e., when \( \alpha /\Gamma \) gets smaller
for fixed \( \xi \)). In the examples shown here, \( \alpha /\Gamma \)
takes on a maximum value of about \( 0.1 \).

\subsection{Optimization of separation quality}

\begin{figure}
\vbox{
\centerline{\psfig{figure=fig6a.eps,width=8cm}}
\centerline{\psfig{figure=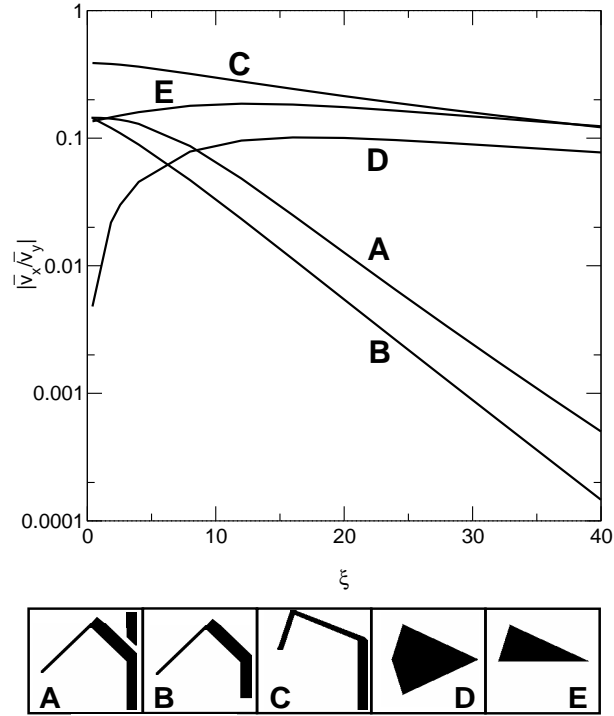,width=8cm}}
}
\caption{\label{deflection} Slope \( \left\langle
x\right\rangle /H=\bar{v}_{x}/\bar{v}_{y} \) plotted
vs. \( \xi \). The obstacle 
shapes are shown at the bottom of the plot and are
discussed in the text. ``A'' refers to the
obstacle shape shown in Fig. \ref{densitycurrent}.
}
\end{figure}

\begin{figure}
\centerline{\psfig{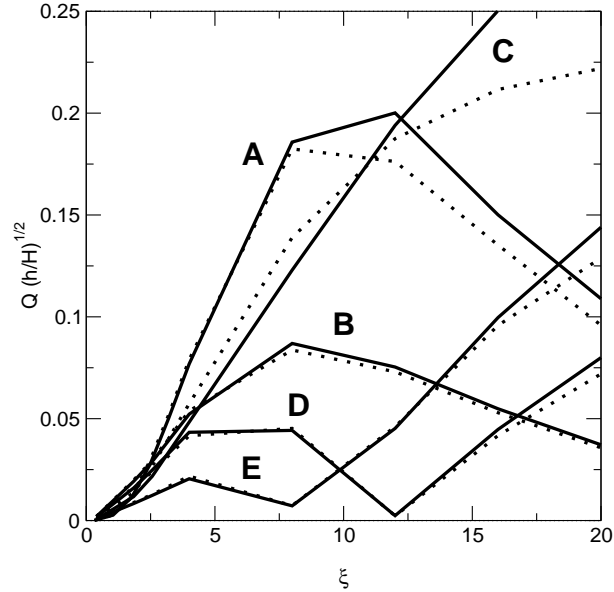}}

\caption{\label{qualityplot}Normalized separation quality \(
Q\sqrt{h/H} \) vs. \( \xi \), where \( \xi \) corresponds to the force
on the first species and \( \xi _{2}=\lambda \xi \), with \( \lambda
=2 \) in this plot. Full lines: results of Monte-Carlo simulation,
dotted lines: Master equation. The difference is due to the larger
values for the diffusion tensor component \( \bar{D}_{xx} \) at high
force \( \xi \) yielded by the master equation. The location of the
optimum does not change much. See discussion in the text.}
\end{figure}

For the obstacle discussed above, the slope \( \bar{v}_{x}/\bar{v}_{y}
\) (proportional to the average deflection \( \left\langle
x\right\rangle \) in the last row) is shown in Fig.~\ref{deflection},
together with those of other obstacle shapes discussed further
below. The exponential decay is consistent with the analytical
estimate derived in a previous section, see Eq.~(\ref{anest}). From
the slope of the logarithmic plot, a value of about \( 0.17 \) has
been obtained for the prefactor \( w^{2}/(4hh') \) in the exponent of
Eq.~(\ref{anest}), which is roughly consistent with the geometrical
parameters of the obstacle. Given the slope and the spread \( \sigma
\) (derived from the components of \( \bar{D} \)), one can obtain the
separation quality \( Q \) defined in Eq.~(\ref{sepquality}), if one
assumes some ratio \( \lambda =\xi _{2}/\xi _{1} \) of the forces
acting on the two species (see Fig.~\ref{qualityplot}).

\subsection{Influence of the obstacle shape}

Both the magnitude of the ratchet effect and the separation quality
depend very much on the shape of the obstacle. We have not performed a
systematic search over obstacle shapes for a kind of ``global''
optimization of the separation quality because of the numerical effort
involved. However, there are a few general properties resulting from
certain geometrical features. These are illustrated by the numerical
results for the slope \( \bar{v}_{x}/\bar{v}_{y} \) and the
separation quality \( Q \) plotted in Figs.~\ref{deflection}
and \ref{qualityplot}. They can be summarized as follows:

The vertical ``wall'' at one side of the obstacle ``A'' depicted in
Fig.~\ref{densitycurrent} acts to prevent particles from diffusing
back to the left, thereby increasing \( \bar{v}_{x} \) and leading to
a better ratchet and separation effect. This can be seen by comparing
against a version with a shorter wall (``B''). The triangular ``roof'' of the
obstacle splits the particle distribution in two halves as it drifts
downwards. If the external force is high, the particles do not have
time to diffuse a sufficient distance to the right and will be
deflected back by the left side of this ``roof'', therefore streaming
downward, with no net deflection to the right. The horizontal position
of the upper tip determines the strength of the force where this
transition takes place: If it is moved to the left (``C''), a much higher
force is necessary. In that case, the slope \( \bar{v}_{x}/\bar{v}_{y}
\) falls off more slowly with increasing force \( \xi \). At the same
time, the value of the slope is generally increased at small forces,
since more particles are deflected one cell to the right.

In order to illustrate the difference between having a strong ratchet
effect and a good separation effect, we have tried a triangular
obstacle (``E''), which yields a comparatively large slope that, however, does
not change very strongly with applied force. By flipping the triangle
along the horizontal axis, an obstacle with reflection symmetry is
created (``D''). This has the peculiar feature \cite{kosterlong,kosturshort}
that, for symmetry reasons, \( \bar{v}_{x} \) is an even function of
the microscopic velocity \( v_{y} \) (driving force \( \xi \)), so the
linear mobility at low driving force vanishes. In principle, this
nonlinear dependence of \( \bar{v}_{x} \) on \( \xi \) is well-suited
for achieving a separation effect. However, it must be kept in mind
that at low values of the external force the diffusive motion is
dominant, so the spread and the overlap of the particle distributions
of the two species in the final row is significant. Therefore, the
separation quality decreases towards \( \xi =0 \).

The separation quality can become zero for a special value of the external
force whenever the slope shown in Fig. \ref{deflection} has an extremum
as a function of microscopic velocity $\xi$, such that two different $\xi$ can
produce the same slope. This occurs for two of the obstacles (D and E in Fig. \ref{qualityplot}).

For most of the obstacle shapes considered here, there is at most a
local maximum of the separation quality at low forces. The global
maximum is expected to occur at much higher values of the force, which
may be unattainable in the experiment (and are difficult to reach in a
numerical simulation with a finite lattice resolution). However, for a
well-suited obstacle like the one depicted in 
Fig.~\ref{densitycurrent}, the quality peaks at moderate values of the
force.

\section{Conclusions}

In this work, we have analyzed a geometric ratchet consisting of a
two-dimensional array of obstacles, where particles perform
drift-diffusive motion under the action of a constant external
force. We have carried out numerical calculations using both a
Monte-Carlo simulation and a master equation solution in order to
obtain the dependence of the ``macroscopic'' drift velocity and
diffusion tensor on external force and obstacle shape. Using these
results, we have quantified the quality of the separation effect that
can be achieved when two species of particles with differing
microscopic mobilities are injected into the array. Our results show
the strong dependence on several features of the shape of the
obstacles and demonstrate the distinction between a strong ratchet
effect and a good separation effect.

\section{Acknowledgments}

We thank Hanno Gassmann for helpful discussions.

\section{Appendix: Calculating the macroscopic diffusion tensor using the 
master equation}

Solving the master equation for the probability density inside a cell
containing a single obstacle allows one to obtain easily a numerically
exact result for the macroscopic drift velocity \( \bar{\vec{v}}
\). It is given by the average flow velocity of particles inside the
cell, i.e., the integral (or sum) over the current density (see
Eqs.~(\ref{driftcurrent}) and (\ref{driftcurrent2})). However, to assess
the quality of separation of a given geometric ratchet, it is equally
important to know the macroscopic diffusion tensor, which governs the
spreading of the macroscopic particle density (i.e., the density
averaged over many obstacles). Its evaluation using the Monte-Carlo
simulation requires a large number of samples, to obtain a good
statistical accuracy. Therefore, it is desirable to calculate \(
\bar{D} \) using the master equation as well. The steps involved in
the derivation of \( \bar{D} \) are slightly more involved than the
straightforward calculation of \( \bar{\vec{v}} \).

Our strategy is to derive the equation of motion for the 
macroscopic density \( \bar{\rho } \), 

\begin{equation}
\label{rhobareq}
\partial _{t}\bar{\rho }=\bar{D}_{ij}\partial _{i}\partial
_{j}\bar{\rho }-
\bar{\vec{v}}\vec{\nabla }\bar{\rho }\, ,
\end{equation}
from the analogous equation for the microscopic density \( \rho \),
making use of the slow variation of \( \bar{\rho } \).

If the particle density \( \rho \) is spread over many obstacle cells
(as is the case after waiting for a sufficiently long time), it may,
in a first approximation, be described by

\begin{equation}
\label{rhorough}
\rho =\rho _{0}\bar{\rho }\, ,
\end{equation}

where \( \rho _{0} \) refers to the detailed density that varies on
the scale of a single obstacle but is periodic throughout the
array. \( \rho _{0} \) has been obtained before by solving the master
equation for a single cell, using periodic boundary conditions at the
borders of the cell and the restriction for the current density \(
\vec{j}_{0} \) to run parallel to the walls of the obstacle. If \(
\bar{\rho } \) were constant, this would constitute a
(not normalizable) stationary periodic solution to the Fokker-Planck
equation. However, \( \bar{\rho } \) is assumed to vary (very
slowly), such that this is not a stationary solution and does not
fulfill the boundary conditions exactly. Therefore, it has to be
supplemented by ``correction terms'', which depend (necessarily
linearly) on the spatial derivatives of \( \bar{\rho }
\). Consequently, we take \( \rho \) to be given by:

\begin{equation}
\label{rhodef}
\rho =\rho _{0}\bar{\rho }+g^{i}\partial _{i}\bar{\rho }+
K^{ij}\partial _{i}\partial _{j}\bar{\rho }+\ldots \, ,
\end{equation}

Here, \( g^{i} \) and \( K^{ij} \) are as yet unknown periodic
functions (that vary on the scale of a single obstacle). We emphasize
that, of course, many different microscopic densities \( \rho \) yield
the same macroscopic density \( \bar{\rho } \). Therefore, \( \rho \)
is not uniquely specified if \( \bar{\rho } \) is given. However, in
the long-time limit assumed here, \( \rho \) has ``equilibrated'' and
the deviations from \( \rho _{0} \) are in a one-to-one correspondence
with the lowest order spatial derivatives of \( \bar{\rho } \).

Our further strategy is as follows: We will first rederive the known
result for the macroscopic drift velocity \( \bar{\vec{v}} \), which
is the constant coefficient appearing in the part of the equation of
motion for \( \bar{\rho } \) that contains the spatial derivatives of
first order: \( \partial _{t}\bar{\rho }=-\bar{\vec{v}}\vec{\nabla
}\bar{\rho } \). To this end, we will insert the Ansatz
(\ref{rhodef}) into the equation of motion for \( \rho \), keeping
only the terms containing up to first derivatives of \( \bar{\rho } \)
and eliminating \( g^{i} \) using the boundary condition. Using the
result for \( \bar{\vec{v}} \), we will obtain an inhomogeneous linear
equation for \( g^{i} \) itself, which must be solved numerically. The
relation between \( g^{i} \) and the macroscopic diffusion tensor \(
\bar{D} \) will be obtained by going to the second order in the
spatial derivatives of \( \bar{\rho } \). Although this involves the
unknown function \( K^{ij} \), we will be able to eliminate it in the
same way that \( g^{i} \) had been eliminated in the first step.

Let us first derive a boundary condition for \( g^{i} \). The current
densities for the first and second term on the right hand side of
Eq.~(\ref{rhodef}) are given by

\begin{eqnarray}
\vec{j} & = & \vec{j}_{0}\bar{\rho }-D\rho _{0}
\left( \vec{\nabla }\bar{\rho }\right) \nonumber \\
(\vec{j}_{1})_{l} & = & -D(\partial _{l}g^{i})(\partial _{i}
\bar{\rho})+\vec{v}_{l}g^{i}(\partial _{i}\bar{\rho })-
Dg^{i}\partial _{l}\partial _{i}\bar{\rho }\, , \nonumber\\
\label{jjonedef}
\end{eqnarray}

Here, \( \vec{j}_{0}=(-D\vec{\nabla }+\vec{v})\rho _{0} \) is the
current density of \( \rho _{0} \) alone. We demand \(
\vec{j}+\vec{j}_{1} \) to be parallel to the obstacle wall, keeping
only terms including first order derivatives of \( \bar{\rho } \) and
then canceling these terms. This leads to the following boundary
condition for \( g^{i} \):

\begin{equation}
\label{boundgi}
\hat{n}\vec{j}_{g^{i}}=D\hat{n}_{i}\rho _{0}\, ,
\end{equation}

where \( \vec{j}_{g^{i}}=(-D\vec{\nabla }+\vec{v})g^{i} \) is the current
density related to \( g^{i} \) and \( \hat{n} \) is the outer normal
vector of the obstacle wall.

In the next step, we calculate \( \partial _{t}\rho =-\vec{\nabla
}(\vec{j}+\vec{j}_{1}+\ldots ) \) up to first order in the spatial
derivatives of \( \bar{\rho } \) and demand it to equal \( \rho
_{0}\partial _{t}\bar{\rho }\approx -\rho _{0}\bar{\vec{v}}\vec{\nabla
}\bar{\rho } \), which is essentially the drift term for \( \bar{\rho
} \). Physically, this equation means that a nonvanishing slope of \(
\bar{\rho } \) will lead to an overall increase (or decrease) of the
microscopic density \( \rho \) inside an obstacle cell. The detailed
shape of the distribution within that cell is not changed, only its
magnitude. After dropping the overall factor \( \partial _{i}\bar{\rho
} \), we arrive at

\begin{equation}
\label{geq}
\vec{\nabla }\vec{j}_{g^{i}}=-2(\vec{j}_{0})_{i}+(\vec{v}_{i}+
\bar{\vec{v}}_{i})\rho _{0}\, .
\end{equation}

It is not necessary to know \( g^{i} \) in order to obtain \(
\bar{\vec{v}} \). We integrate both sides of this equation over the
whole cell, assuming periodic boundary conditions for \( g^{i} \) (as
well as for \( \rho _{0} \)). The boundary term resulting from the
walls of the obstacle contains \( g^{i} \), but it can be transformed
using Eq.~(\ref{boundgi}), such that we end up with an equation where
\( g^{i} \) has been eliminated:

\begin{equation}
\label{resvbar}
-D\oint \hat{n}_{i}\rho _{0}ds=-2\int (\vec{j}_{0})_{i}d^{2}\vec{r}+
(\vec{v}_{i}+\bar{\vec{v}}_{i})\, . 
\end{equation}

The integral on the left-hand side runs along the obstacle wall, while
that on the right hand side extends over the whole cell. Since
\begin{equation}
\label{joint}
\int (\vec{j}_{0})_{i}d^{2}\vec{r}=\vec{v}_{i}+D\oint 
\hat{n}_{i}\rho _{0}ds\, ,
\end{equation}
we have

\begin{equation}
\label{vbar}
\bar{\vec{v}}_{i}=\int (\vec{j}_{0})_{i}d^{2}\vec{r}\, ,
\end{equation}
as had been assumed in the main text already 
(see Eqs.~(\ref{driftcurrent}) and (\ref{driftcurrent2})). 

Inserting \( \bar{\vec{v}} \) into Eq.~(\ref{geq}) yields an
inhomogeneous linear partial differential equation for \( g^{i} \)
which has to be solved numerically (assuming periodicity and the
boundary condition Eq.~(\ref{boundgi})). Note that \( g^{i} \) is only
determined up to a constant multiple of \( \rho _{0} \), since \( \rho
_{0} \) solves the homogeneous equation. However, as we will see, this
does not affect the result for the diffusion tensor \( \bar{D} \) to
be derived from \( g^{i} \). Further remarks concerning the numerical
solution of the master equation on the discrete lattice can be found
at the end of this appendix.

The current density related to the part of \( \rho \) that involves second
derivatives of \( \bar{\rho } \) (see Eq.~(\ref{rhodef})) is equal to

\begin{equation}
\label{jtwodef}
(\vec{j}_{2})_{l}=(\partial _{i}\partial _{j}\bar{\rho })
\left[ (-D\vec{\nabla }_{l}+\vec{v}_{l})K^{ij}\right] +\ldots \, ,
\end{equation}

where we have neglected higher derivatives of \( \bar{\rho } \). We
arrive at a boundary condition for \( K^{ij} \) at the walls of the
obstacle in the same way as for \( g^{i} \), by demanding \(
\vec{j}_{tot}\equiv \vec{j}+\vec{j}_{1}+\vec{j}_{2} \) to be parallel
to the wall. This time, we keep only the terms including second
derivatives of \( \bar{\rho } \). This leads to

\begin{equation}
\label{Kbound}
\left[ \hat{n}\left( -D\vec{\nabla }+\vec{v}\right) \right] 
K^{ij}=D\hat{n}_{j}g^{i}\, .
\end{equation}

To this order, the time-derivative of \( \rho \), \( \partial _{t}\rho
=-\vec{\nabla }\vec{j}_{tot} \), includes both the diffusion of \(
\bar{\rho } \) and the drift of the term \( g^{i}\partial
_{i}\bar{\rho } \):

\begin{equation}
\label{divjtot}
\vec{\nabla }\vec{j}_{tot}=
\bar{\vec{v}}_{l}g^{i}\partial _{l}\partial _{i}\bar{\rho }
-\rho _{0}\bar{D}_{li}\partial _{l}\partial _{i}\bar{\rho }+\ldots\, ,
\end{equation}

keeping only the second order with respect to the spatial derivatives
of \( \bar{\rho } \) on both sides of the equation.

As before, we integrate this equation over the cell and use the
boundary condition Eq.~(\ref{Kbound}) at the obstacle walls to
eliminate \( K^{ij} \). The resulting expression for \( \bar{D} \)
then is given by:

\begin{equation}
\label{dbar}
\bar{D}_{ji}=D\left( \delta _{ji}-\oint \hat{n}_{j}g^{i}ds\right) + 
(\bar{\vec{v}}_{j}-\vec{v}_{j})\int g^{i}d^{2}\vec{r}\, .
\end{equation}

Note that adding \( \lambda \rho _{0} \) to \( g^{i} \) (with an arbitrary
constant \( \lambda \)) does not affect the result for \( \bar{D} \), due
to Eqs.~(\ref{joint}) and (\ref{vbar}).

For the numerical solution, it is, in principle, possible to
discretize the equations (\ref{boundgi}) and (\ref{geq}) for \( g^{i}
\) as well as the expression (\ref{dbar}) for \( \bar{D} \). However,
this is guaranteed to coincide with the results of the Monte-Carlo
simulation only in the continuum limit (where, e.g., \( \alpha /\Gamma
\rightarrow 0 \)). In order to have a better agreement even when one
is not yet in the continuum limit, it is advisable to start directly
from the discretized master equation and redo the steps of the
derivation shown here for the discrete lattice.

The equation which has actually been solved numerically to arrive at
\( g^{i} \) is given by:

\begin{equation}
\label{numdiffmeq}
Lg^{i}=-p\bar{\vec{v}}_{i}-\Gamma 
(p^{i+}-p^{i-})-\delta _{i2}\alpha (p^{i+}+p^{i-})\, .
\end{equation}

Here, \( Lg^{i} \) corresponds to \( -\vec{\nabla }\vec{j}_{g^{i}}
\). Therefore, \( L \) is the matrix kernel which is also used for
solving the homogeneous equation, \( Lp=0 \), including the same
treatment of obstacle walls and periodic boundary conditions (see the
right-hand side of Eq.~(\ref{Ldef}) for the definition of \( L \)). \(
p \) is evaluated at the ``current site'' (the site which the left
hand side refers to), while \( p^{i+} \) and \( p^{i-} \) are
evaluated at the neighboring sites, in positive or negative direction
\( i \) (=\( 1,2 \), corresponding to \( x,y \)), respectively. At
obstacle walls, these neighboring sites may turn out to be
``forbidden'', in which case \( p^{i+} \) or \( p^{i-} \)
vanishes. This implements the discrete version of the boundary
condition discussed above. (Note that according to the convention used
here, the microscopic drift velocity is assumed to point in negative
\( y \)-direction if \( \alpha \) is positive).

In order to evaluate \( \bar{D} \), we must carry out a sum over all sites
at the wall of the obstacle, i.e., those which have forbidden sites as 
neighbors.
This sum is denoted by \( \sum _{W} \). The sum extending over all allowed
sites in the cell is denoted by \( \sum \):

\begin{equation}
\label{difftensordiscrete}
\bar{D}_{ji}=\Gamma \left( \delta _{ji}-\sum_{W}\hat{n}_{j}g^{i}\right)
+(\bar{\vec{v}}_{j}-\vec{v}_{j})
\sum g^{i}-\alpha \delta _{j2}\sum _{W}g^{i}\, .
\end{equation}

The last term vanishes in the continuum limit but is important to ensure that
\( \bar{D} \) does not change on adding a homogeneous solution 
\( \lambda p \) to \( g^{i} \).


\begin{thebibliography}{10}

\bibitem{reviews} P. Reimann, cond-mat/0010237, to appear in Physics Reports.

\bibitem{astumianmolmotor} R.~D. Astumian and M. Bier,
Phys. Rev. Lett. {\bf 72}, 1766 (1994).

\bibitem{brownianmotorscience} R.~D. Astumian, Science {\bf 276}, 917 (1997).

\bibitem{fluxonpump} J.~F. Wambaugh {\it et al.}, 
Phys. Rev. Lett. {\bf 83}, 5106 (1999).

\bibitem{faucheux} L.~P. Faucheux, L.~S. Bourdieu, P.~D. Kaplan, 
and A.~J. Libchaber, 
Phys. Rev. Lett. {\bf 74}, 1504 (1995).

\bibitem{haenggi} P. H\"anggi and R. Bartussek, in: 
\emph{Nonlinear Physics of Complex Systems
- Current Status and Future Trends}, edited by J. Parisi, S.~C. M\"uller, and
W. Zimmermann, Springer Lecture Notes in Physics, Vol. {\bf 476}, p. 294
(Springer, Berlin, 1996).

\bibitem{duke} T.~A. Duke and R.~H. Austin, 
Phys. Rev. Lett. {\bf 80}, 1552 (1998).

\bibitem{ertas} D. Ertas, Phys. Rev. Lett. {\bf 80}, 1548 (1998).

\bibitem{derenyi} I. Der\'enyi and R.~D. Astumian, 
Phys. Rev. E {\bf 58}, 7781 (1998).

\bibitem{oudenaarden} A.~v. Oudenaarden and S.~G. Boxer, 
Science {\bf 285}, 1046 (1999).

\bibitem{kosterlong} M. Kostur and L. Schimansky-Geier, 
Phys. Lett. A {\bf 265}, 337 (2000).

\bibitem{kosturshort} M. Bier, M. Kostur, I. Der\'enyi, and
R.~D. Astumian, Phys. Rev. E {\bf 61}, 7184 (2000).

\bibitem{lapack} E. Anderson {\it et al.}, {\it LAPACK Users' Guide}, 
third edition (
SIAM,Philadelphia, 1999).

\end{thebibliography}
\end{document}